\documentclass[journal]{IEEEtran}
\usepackage{color,soul} % ADDED FOR HIGHLIGHTING CHANGES
\usepackage{cite}
\ifCLASSINFOpdf
  \usepackage[pdftex]{graphicx}
  \DeclareGraphicsExtensions{.pdf,.jpeg,.png}
\else
\fi
\usepackage{amsmath}
\ifCLASSOPTIONcompsoc
  \usepackage[caption=false,font=normalsize,labelfont=sf,textfont=sf]{subfig}
\else
  \usepackage[caption=false,font=footnotesize]{subfig}
\fi
\begin{document}
%
% paper title
% Titles are generally capitalized except for words such as a, an, and, as,
% at, but, by, for, in, nor, of, on, or, the, to and up, which are usually
% not capitalized unless they are the first or last word of the title.
% Linebreaks \\ can be used within to get better formatting as desired.
% Do not put math or special symbols in the title.
\title{Clutter distributions for tomographic image standardization in ground-penetrating radar}
%
%
% author names and IEEE memberships
% note positions of commas and nonbreaking spaces ( ~ ) LaTeX will not break
% a structure at a ~ so this keeps an author's name from being broken across
% two lines.
% use \thanks{} to gain access to the first footnote area
% a separate \thanks must be used for each paragraph as LaTeX2e's \thanks
% was not built to handle multiple paragraphs
%
\author{Brian~M.~Worthmann,
        David~H.~Chambers,~\IEEEmembership{Senior~Member,~IEEE,}
        David~S.~Perlmutter,
        Jeffrey~E.~Mast,
        David~W.~Paglieroni,~\IEEEmembership{Senior~Member,~IEEE,}
        Christian~T.~Pechard,
        Garrett~A.~Stevenson,
        and~Steven~W.~Bond,~\IEEEmembership{Member,~IEEE}%
        \thanks{The authors would like to acknowledge support from the Office of Naval Research. This work was performed under the auspices of the U.S. Department of Energy by Lawrence Livermore National Laboratory under Contract DE-AC52-07NA27344.}%
        \thanks{B.M.~Worthmann, D.H.~Chambers, D.S.~Perlmutter, D.W.~Paglieroni, C.T.~Pechard, G.A.~Stevenson, and S.W.~Bond are with Lawrence Livermore National Laboratory, Livermore, CA 94550 USA. (e-mails, respectively: worthmann1@llnl.gov, chambers2@llnl.gov, perlmutter1@llnl.gov, paglieroni1@llnl.gov, pechard1@llnl.gov, stevenson32@llnl.gov and bond6@llnl.gov).}%
        \thanks{Jeffrey E. Mast is with Teres Technologies, Inc., Loveland, CO 80537 USA (e-mail: jeff@terestech.com).}
        \thanks{Manuscript submitted 3 Aug 2020}}

% note the % following the last \IEEEmembership and also \thanks - 
% these prevent an unwanted space from occurring between the last author name
% and the end of the author line. i.e., if you had this:
% 
% \author{....lastname \thanks{...} \thanks{...} }
%                     ^------------^------------^----Do not want these spaces!
%
% a space would be appended to the last name and could cause every name on that
% line to be shifted left slightly. This is one of those "LaTeX things". For
% instance, "\textbf{A} \textbf{B}" will typeset as "A B" not "AB". To get
% "AB" then you have to do: "\textbf{A}\textbf{B}"
% \thanks is no different in this regard, so shield the last } of each \thanks
% that ends a line with a % and do not let a space in before the next \thanks.
% Spaces after \IEEEmembership other than the last one are OK (and needed) as
% you are supposed to have spaces between the names. For what it is worth,
% this is a minor point as most people would not even notice if the said evil
% space somehow managed to creep in.

% The paper headers
\markboth{Journal of \LaTeX\ Class Files,~Vol.~14, No.~8, August~2015}%
{Shell \MakeLowercase{\textit{et al.}}: Bare Demo of IEEEtran.cls for IEEE Journals}
% The only time the second header will appear is for the odd numbered pages
% after the title page when using the twoside option.
% 
% *** Note that you probably will NOT want to include the author's ***
% *** name in the headers of peer review papers.                   ***
% You can use \ifCLASSOPTIONpeerreview for conditional compilation here if
% you desire.

% If you want to put a publisher's ID mark on the page you can do it like
% this:
%\IEEEpubid{0000--0000/00\$00.00~\copyright~2015 IEEE}
% Remember, if you use this you must call \IEEEpubidadjcol in the second
% column for its text to clear the IEEEpubid mark.

% use for special paper notices
%\IEEEspecialpapernotice{(Invited Paper)}

% make the title area
\maketitle

% As a general rule, do not put math, special symbols or citations
% in the abstract or keywords.
\begin{abstract}
Multistatic ground-penetrating radar (GPR) signals can be imaged tomographically to produce three-dimensional distributions of image intensities. In absence of objects of interest, these intensities can be considered to be estimates of clutter. These clutter intensities spatially vary over several orders of magnitude, and vary across different arrays, which makes direct comparison of these raw intensities difficult. However, by gathering statistics on these intensities and their spatial variation, a variety of metrics can be determined. In this study, the clutter distribution is found to fit better to a two-parameter Weibull distribution than Gaussian or lognormal distributions. Based upon the spatial variation of the two Weibull parameters, scale and shape, more information may be gleaned from these data. How well the GPR array is illuminating various parts of the ground, in depth and cross-track, may be determined from the spatial variation of the Weibull scale parameter, which may in turn be used to estimate an effective attenuation coefficient in the soil. The transition in depth from clutter-limited to noise-limited conditions (which is one possible definition of GPR penetration depth) can be estimated from the spatial variation of the Weibull shape parameter. Finally, the underlying clutter distributions also provide an opportunity to standardize image intensities to determine when a statistically significant deviation from background (clutter) has occurred, which is convenient for buried threat detection algorithm development which needs to be robust across multiple different arrays.
\end{abstract}

% Note that keywords are not normally used for peerreview papers.

% For peer review papers, you can put extra information on the cover
% page as needed:
% \ifCLASSOPTIONpeerreview
% \begin{center} \bfseries EDICS Category: 3-BBND \end{center}
% \fi
%
% For peerreview papers, this IEEEtran command inserts a page break and
% creates the second title. It will be ignored for other modes.
\IEEEpeerreviewmaketitle

\section{Introduction}
\label{Sec1}
% The very first letter is a 2 line initial drop letter followed
% by the rest of the first word in caps.
% 
% form to use if the first word consists of a single letter:
% \IEEEPARstart{A}{demo} file is ....
% 
% form to use if you need the single drop letter followed by
% normal text (unknown if ever used by the IEEE):
% \IEEEPARstart{A}{}demo file is ....
% 
% Some journals put the first two words in caps:
% \IEEEPARstart{T}{his demo} file is ....
% 
% Here we have the typical use of a "T" for an initial drop letter
% and "HIS" in caps to complete the first word.

\IEEEPARstart{G}{round}-penetrating radar (GPR) allows for buried object detection, such as landmines or utility pipes. GPR can be used to detect spatial changes in dielectric permittivity, meaning both metallic and nonmetallic objects are detectable. In the system used for this study, a vehicle-mounted multistatic radar array takes time series measurements, which are used to form three dimensional images in real-time. In the absence of buried objects, the images are not empty, but rather contain clutter, where the word ``clutter'' is used here to denote random volumetric scattering associated with spatial variations in soil permittivity -- in other words, homogeneous scattering associated with soil composition and structure, as opposed to scattering associated with discrete buried objects such as rocks or man-made objects. The amplitudes of those clutter voxels are then statistically analyzed to determine their deviation from the background. This study discusses the statistics of those clutter voxels, including empirical fits to probability distribution functions (PDFs), spatial variations of the distributions' parameters, and how standardization can be performed on image voxels to improve buried object visibility.

\subsection{Clutter statistics}
 Most literature involving the fitting of clutter amplitudes or intensities to probability distribution functions (PDFs) is associated with monostatic radars looking at weather-, land- or sea-based clutter, with the latter receiving the most attention. A variety of clutter distributions are considered. These include one-parameter distributions, such as the zero-mean Gaussian model \cite{Ref1, Ref2} which was an adequate model for thermal noise, and the Rayleigh distribution for amplitudes \cite{Ref3} (or chi-squared distribution for intensities \cite{Ref4}) which was found to be suitable for high-grazing angle radar turns. A number of two-parameter distributions have been proposed for handling low-grazing angle returns, including the lognormal distribution \cite{Ref3}, the Weibull distribution \cite{Ref5, Ref6}, the Rice distribution for amplitudes \cite{Ref7} (or noncentral chi-squared distribution for intensities \cite{Ref4}), the heavy-tailed Rayleigh distribution \cite{Ref8}, and the K-distribution \cite{Ref9, Ref10, Ref11} and similar generalizations \cite{Ref12}. For additional flexibility, more parameters may be added, such as with the six-parameter generalized compound distribution \cite{Ref13}, which contains within it numerous special cases, including the generalized gamma distribution, the hypergeometric distribution, and all of the other one- and two-parameter systems named here. The overview provided here is far from complete, as there have been several decades of research into radar clutter modeling \cite{Ref14}. It is also noted that some of these distributions have physical interpretations for their parameters, while others are simply empirical fits. In this study, only the two-parameter lognormal and Weibull distributions are considered, for simplicity. Some physical interpretation of the shape and scale parameters is provided, but the parameters are fundamentally an empirical fit. This is in agreement with the claim by Shnidman that the Weibull and log-normal distributions are used because “they produce a reasonable fit to measured clutter densities, rather than any underlying physical argument” \cite{Ref4}.
 
 There are other investigations into clutter associated with ground penetrating radar, including physical modeling of soil scattering \cite{Ref18}, a graphical method to determine goodness-of-fit for different clutter probability distributions \cite{Ref19}, and a study on the relationships between complex permittivity and the observed signal-to-clutter ratios \cite{Ref20}. These studies all utilize statistics gathered on time-series data directly, whereas the study here utilizes ground penetrating radar \textit{imaging} data, formed from an application of computed tomography on the measured time-series data.
 
 While the GPR system utilized for this study is downward-looking, many characteristics are shared with forward-looking GPR systems. Computed tomography images associated with forward-looking GPR are detailed in \cite{8417956}, \cite{7942118} and \cite{Comite2020}. Several studies (e.g. \cite{4358826}, \cite{885209} and \cite{662655}) use electromagnetic simulations to study the effects of clutter originating from rough surface scattering, and present methods to reduce that type of clutter. A detailed study by Liao and Dogaru \cite{6204052} showed the quantitative effect of rough surface scattering, and determined that scattering amplitudes are well-described by K-distributions, or Rayleigh distributions at low surface roughness. Another forward-looking GPR imaging study \cite{7942118} found clutter amplitudes follow a truncated Rayleigh distribution, and targets follow a three-component Gaussian mixture model. In this study, clutter intensities are empirically found to well-described by lognormal and Weibull distributions instead (see Section \ref{Sec2_1}).
 
 \subsection{Multistatic tomographic imaging for GPR}
\begin{figure}[!t]
\centering
\includegraphics[width=0.75\linewidth]{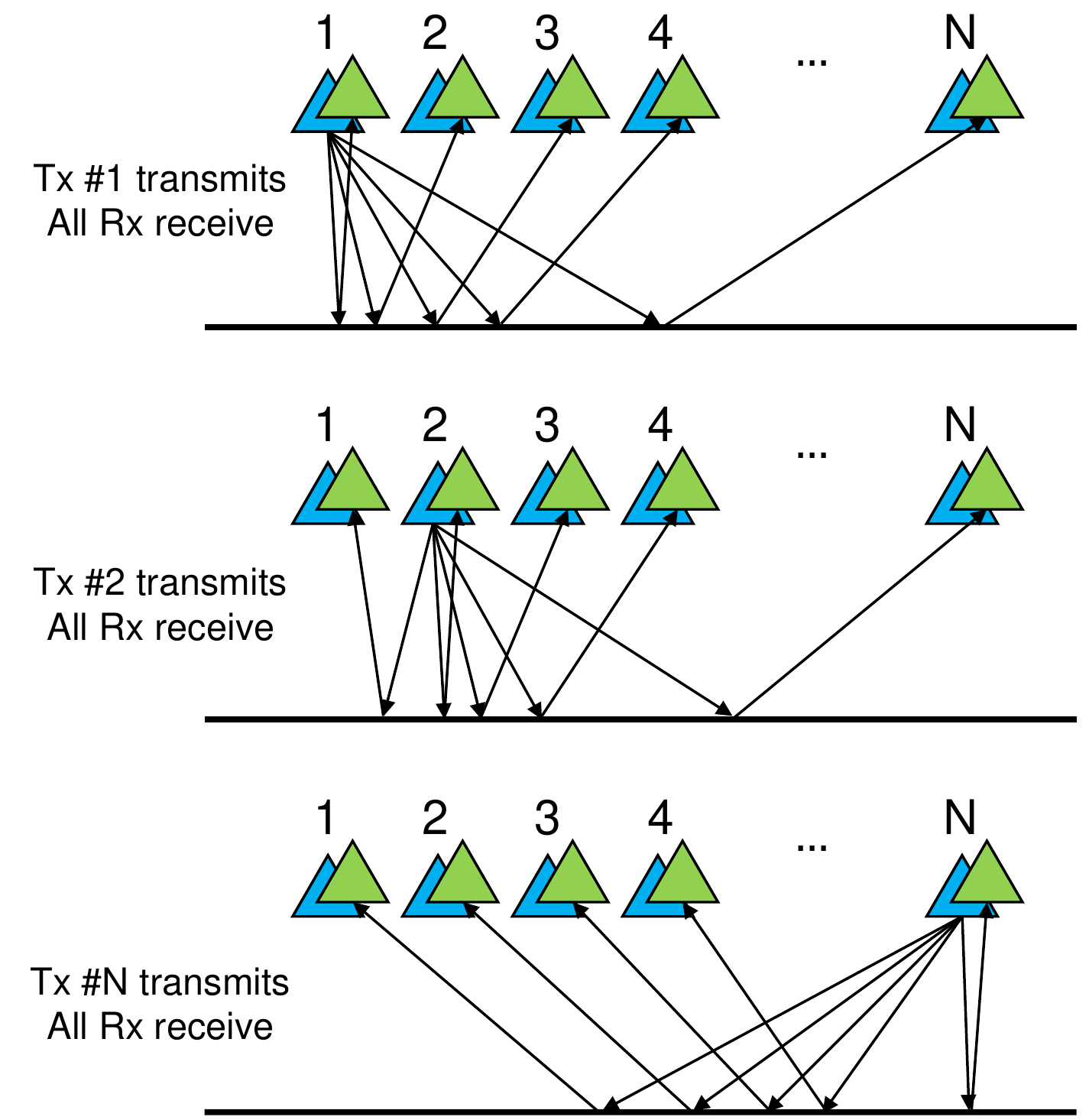}
\caption{Schematic illustrating the multistatic transmitter and receiver architecture. There are $N$ transmitters (in blue) and $N$ receivers (in green). The black horizontal lines represent the ground, and only the arrows represent the reflections from the ground, though for clarity, other reflections, such as those from subsurface objects or clutter are omitted for clarity of the figure.}
\label{Fig1}
\end{figure}

 Many commercially available GPR systems operate in a bistatic or multi-monostatic configuration, where only one transmitter and one receiver are active at a time. In the vehicle-based system \cite{Ref15} used for this study, a multistatic array is used, where only one transmitter (Tx) is active at a time, but all receivers (Rx) are taking measurements, as illustrated in Figure \ref{Fig1}. The array used here has $N=16$ transmitters and an equivalent number of receivers located at approximately the same position in cross-track and are separated by a small distance in down-track. After cycling through all 16 transmitters, $16^2=256$ time series measurements are stored, representing one frame. This is an increase in the amount of received data by a factor of $16$ when compared to a multi-monostatic configuration. These additional measurements, and their associated spatial diversity, permit improved signal-to-noise ratios compared to a multi-monostatic system.
 
\begin{figure}
\centering
\subfloat[]{\includegraphics[width=0.58\linewidth]{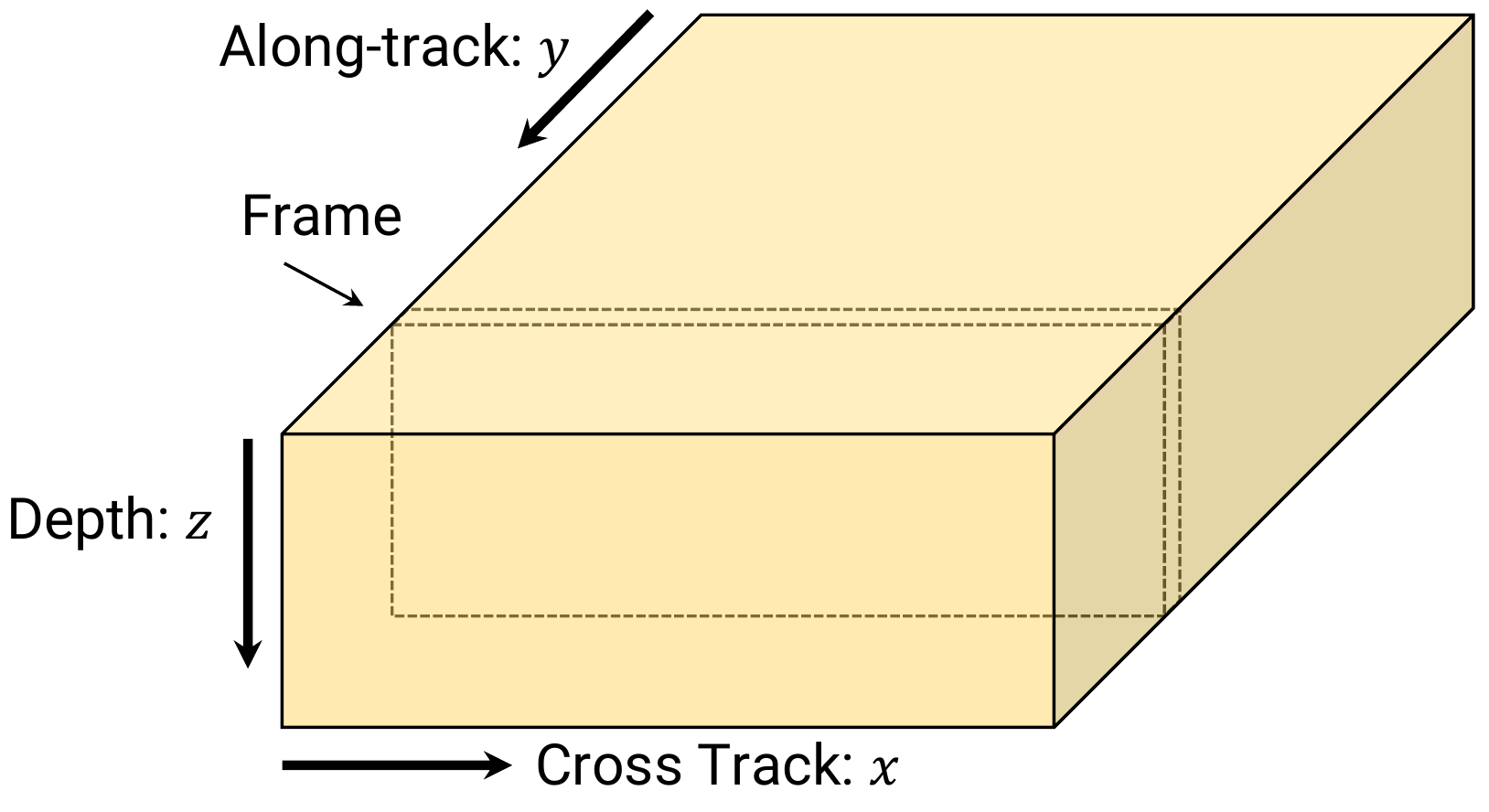}
    \label{Fig2a}}
\subfloat[]{\includegraphics[width=0.38\linewidth]{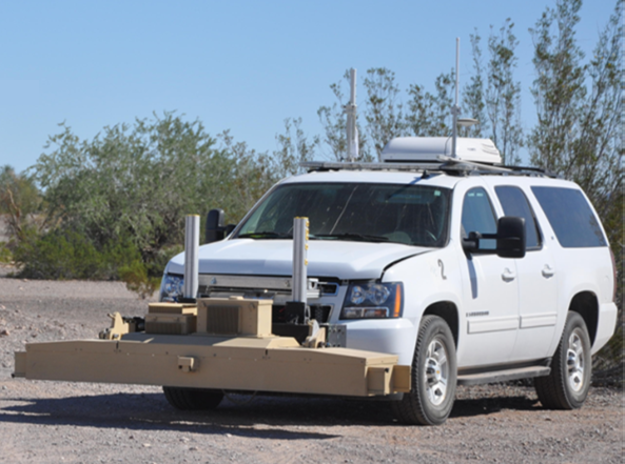}
    \label{Fig2b}}
\caption{Left: Schematic for the spatial convention used here, with $x$ serving as the cross-track variable, $y$ serving as the along-track variable, and $z$ serving as the depth variable (positive oriented into the ground). Right: A photograph of the vehicle with the GPR array mounted to the front of the vehicle. The array is nominally 30~cm above the surface.}
\label{Fig2}
\end{figure}

Each frame’s time series are pre-processed to remove the coupling pulse, radio interference, and the reflection from the ground, leaving only the radar returns from subsurface features, such as buried objects or volumetric scattering (clutter). Note that imperfect removal of the surface reflection would also lead to clutter, but this study does not distinguish between origins of clutter (i.e. surface roughness or volumetric scattering). The pre-processed time series serve as inputs to a plane-to-plane backpropagation algorithm \cite{Ref16}, which leverages temporal and spatial Fourier transforms to focus the received signal amplitudes to a uniformly-spaced spatial grid. The focused tomographic image serves as the data utilized for this study, as well as the input to detection and classification algorithms \cite{Ref15, Ref17}. The spatial convention used here is illustrated in Figure \ref{Fig2}, along with a photograph of the vehicle with the GPR array mounted to the front.

In the construction of the images used in this study, not all of the $256$ time series are utilized. Visualizing the time series as a $16\times16$ matrix of time series, if only the diagonal were used, that would be comparable to a multi-monostatic array. We found empirically that utilizing matrix elements within four rows or columns of the diagonal gave better imaging performance. In other words, using the language of reference \cite{Ref17}, the multistatic degree is set to four (where a multistatic degree of zero corresponds to the multi-monostatic case). This choice is primarily limited by the beamwidth of the antenna, with a wider beamwidth supporting a higher multistatic degree. Given the receivers are spaced at 15~cm, and the nominal array height above the ground is 30~cm, a multistatic degree of four is equivalent to using surface reflections of 45 degrees or less. The antenna used in this study has an approximate half-width half-max beamwidth of 20--25 degrees, which means the far off-diagonal elements receive approximately 12 dB less power (or 25\% of the amplitude) as the mononstatic (diagonal) elements receive.

Since the backpropagation algorithm operates in the frequency domain, only certain frequencies (100 MHz to 3.5 GHz) are imaged in this study. The synthetic aperture radar (SAR) algorithm detailed in Chambers et al. \cite{Ref15} was turned off for the purposes of this study, so that magnitude trends in depth could be assessed without the additional complication of along-track summations that vary in depth. The images that are created are 2.4 meters (m) (128 pixels) in cross-track ($x$), 90cm (46 pixels) in depth ($z$) and a variable number of frames along-track ($y$), with a frame spacing of 2 centimeters (cm). In depth, the first 10 cm correspond to the air gap with a refractive index of 1. The remaining 80 cm correspond to the soil, with an assumed refractive index of 2.0. 

\subsection{Dataset overview}
\begin{figure}[!t]
\centering
\includegraphics[width=\linewidth]{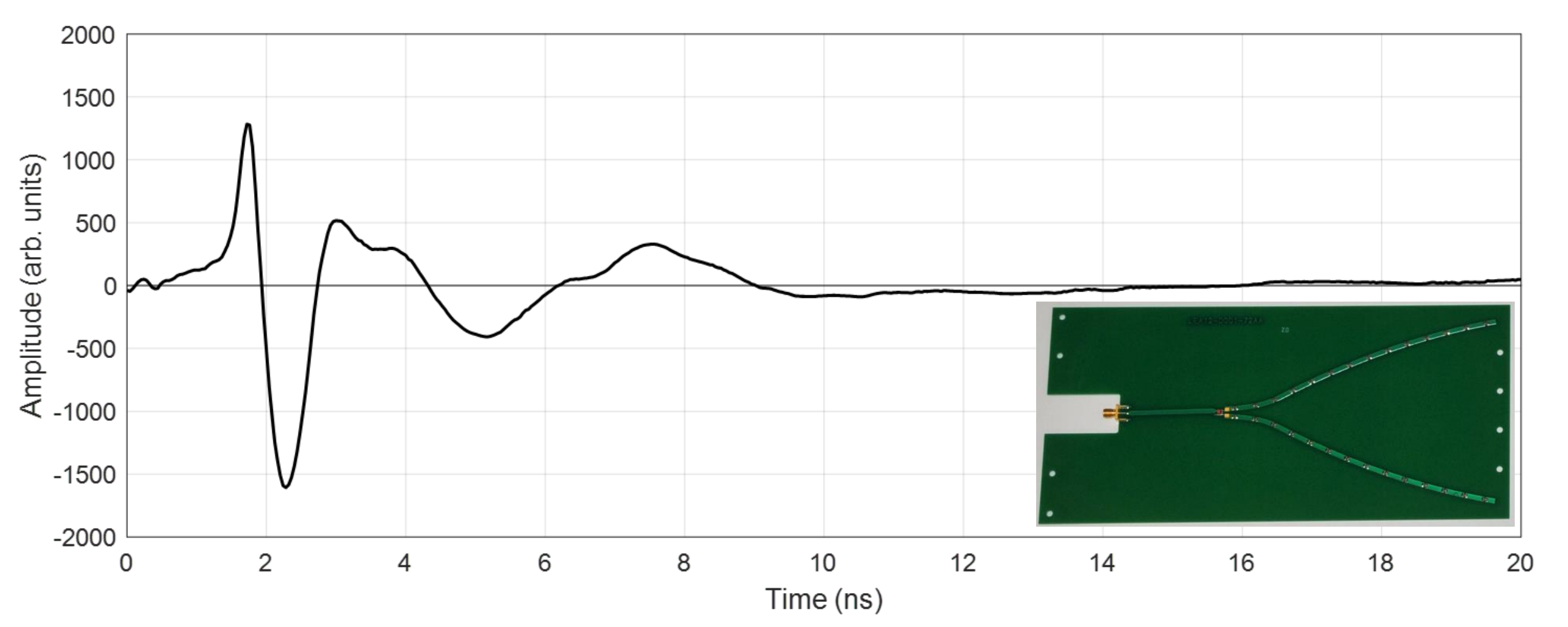}
\caption{Plot of the ultra-wideband pulse shape, amplitude (in arbitrary units) versus time, in nanoseconds. The inset at the bottom-right shows a photograph of the RLVD antenna used.}
\label{Fig3}
\end{figure}

In this study, the data from only one array is used. Specifically, a resistively-loaded V-dipole (RLVD) antenna is used (pictured in Figure \ref{Fig3}), along with its monostatic pulse shape for ground reflections. The center frequency of the pulse is approximately 500 MHz, though all frequencies between 100 MHz and 3.5 GHz are utilized in the plane-to-plane backpropagation imaging algorithm. The same antenna is used for transmission and receiving, meaning there are 32 antennas inside the array pictured in Fig. \ref{Fig2b}. The 16 receiving antennas are spaced 15 cm apart in the cross-track direction for a physical array aperture of 2.25 m. The 16 transmitting antennas have the same spacing, and are offset from the row of receivers by 26.5 cm in the along-track direction. Due to the antenna beamwidth, the imaging aperture of the array is found empirically to be approximately 2.4 m. 

The data used here were taken across two campaigns (separated by one year) to a southwestern United States GPR testing location. A total of 12 lanes at this location were imaged by this array, representing approximately 9.3 km (or 5.5 acres) of unique earth. Each run (a single pass over a single lane) is comprised of thousands of frames—on average, 40,000 frames per run. Each frame represents 2 cm (1 voxel) in the along-track direction, 2.4 m (128 voxels) in the cross-track direction, and 0.9 m (46 voxels) in the depth direction. A total of 189 runs were used in this study, representing 45 billion individual voxels, or approximately 150 km (or 90 acres) of imagery. Due to the spatial sparseness of objects that were intentionally buried in these lanes, the overwhelming majority of this tomographic image data can be considered to be measurements of the background, or clutter. Thus, for purposes of gathering statistics on the clutter, the first and second moments of the distributions are not expected to be influenced by the presence of buried objects, as these represent less than 1\% of the image volume. While these buried objects have little effect on the first and second moments of the distribution, they may cause deviations from the anticipated distribution for image amplitudes above the 99th percentile -- see discussion in Sec \ref{Sec2_4}.

\subsection{Methodology Overview}
For clarity, we summarize below the processing steps applied to the recorded data which serve as the input data for the analyses given in the remaining sections of this paper.

\begin{enumerate}
    \item Fully multistatic ($16\times16$) time series are recorded every 2~cm.
    \item The coupling pulse is removed from the time series.
    \item The surface reflection is found and subtracted from the time series.
    \item The time series are re-aligned to a nominal height of 10cm above the surface.
    \item Plane-to-plane backpropagation is performed on these time series.
    \item The image is formed through the summation across a range of frequencies (given by the bandwidth) and a range of transmitter/receiver pairs (given by the multistatic degree)
    \item Image magnitudes are stored as a function of depth, cross-track, and along-track.
    \item Repeat steps (1) through (7) for each of 189 runs.
\end{enumerate}

In Section \ref{Sec2}, these image magnitudes are fit to probability distributions. In Section \ref{Sec3}, physical interpretations of the probability distribution parameters are provided. In Section \ref{Sec4}, these image magnitudes are standardized to suppress clutter and accentuate buried targets, and Section \ref{Sec5} provides the conclusions drawn from this study.

\section{Probability Distributions}
\label{Sec2}
\subsection{PDF and CDF definitions}
\label{Sec2_1}
The data described in the previous section are fit to two distributions, the lognormal and Weibull distribution. Their definitions are given below:

\begin{equation}
\label{eqn:PDFlognormal}
P_\text{Lognormal}\left(X=x|\mu,\sigma\right)=\frac{1}{\sqrt{2\pi\sigma^2}}\exp\left(-\frac{\left(\ln x-\mu\right)^2}{2\sigma^2}\right)
\end{equation}

\begin{equation}
    \label{eqn:PDFweibull}
    P_\text{Weibull}\left(X=x|\lambda,\kappa\right)=\frac{\kappa}{\lambda}\left(\frac{x}{\lambda}\right)^{\kappa-1}\exp\left(-\left(\frac{x}{\lambda}\right)^\kappa\right)
\end{equation}

Thus, $\mu$ and $\lambda$ can be seen as ``scale'' parameters, while $\sigma$ and $\kappa$ can be seen as ``shape'' parameters. $X$ is the random variable containing the clutter amplitudes, and $x$ is the corresponding dummy variable. The cumulative distribution function of each function is given by:

\begin{equation}
\label{eqn:CDFlognormal}
\text{CDF}_\text{Lognormal}(X=x|\mu,\sigma)=\frac{1}{2}\left(1+\text{erf}\left(\frac{\ln x-\mu}{\sqrt{2}\sigma}\right)\right)
\end{equation}

\begin{equation}
    \label{eqn:CDFweibull}
    \text{CDF}_\text{Weibull}\left(X=x|\lambda,\kappa\right)=1-\exp\left(-\left(\frac{x}{\lambda}\right)^\kappa\right)
\end{equation}

\subsection{Distribution parameter estimates}
To estimate lognormal distribution parameters, the following equations are used, where $E(\cdot)$ represents an expected value:
\begin{equation}
    \mu = E\left(\ln{X}\right)
\end{equation}

\begin{equation}
    \sigma = \sqrt{E\left((\ln{X})^2\right) - \left(E\left(\ln{X}\right)\right)^2}
\end{equation}

To estimate the Weibull distribution parameters, the first and second moments of the clutter amplitude data, $E(X)$ and $E(X^2)$ can be used:

\begin{equation}
    \frac{\Gamma\left(1+\frac{2}{\kappa}\right)}{\Gamma\left(1+\frac{1}{\kappa}\right)^2} = \frac{E(X^2)}{E(X)^2}
    \label{Eqn8}
\end{equation}

\begin{equation}
    \lambda=\frac{E(X)}{\Gamma\left(1+\frac{1}{\kappa}\right)}
\end{equation}

In Eqn. \ref{Eqn8}, it is not possible to analytically evaluate $\kappa$ directly, however the function on the left is well behaved, so a simple numerical interpolation scheme is sufficient to determine $\kappa$, given the right-hand side ratio of moments.

\begin{figure}
\centering
\includegraphics[width=0.85\linewidth]{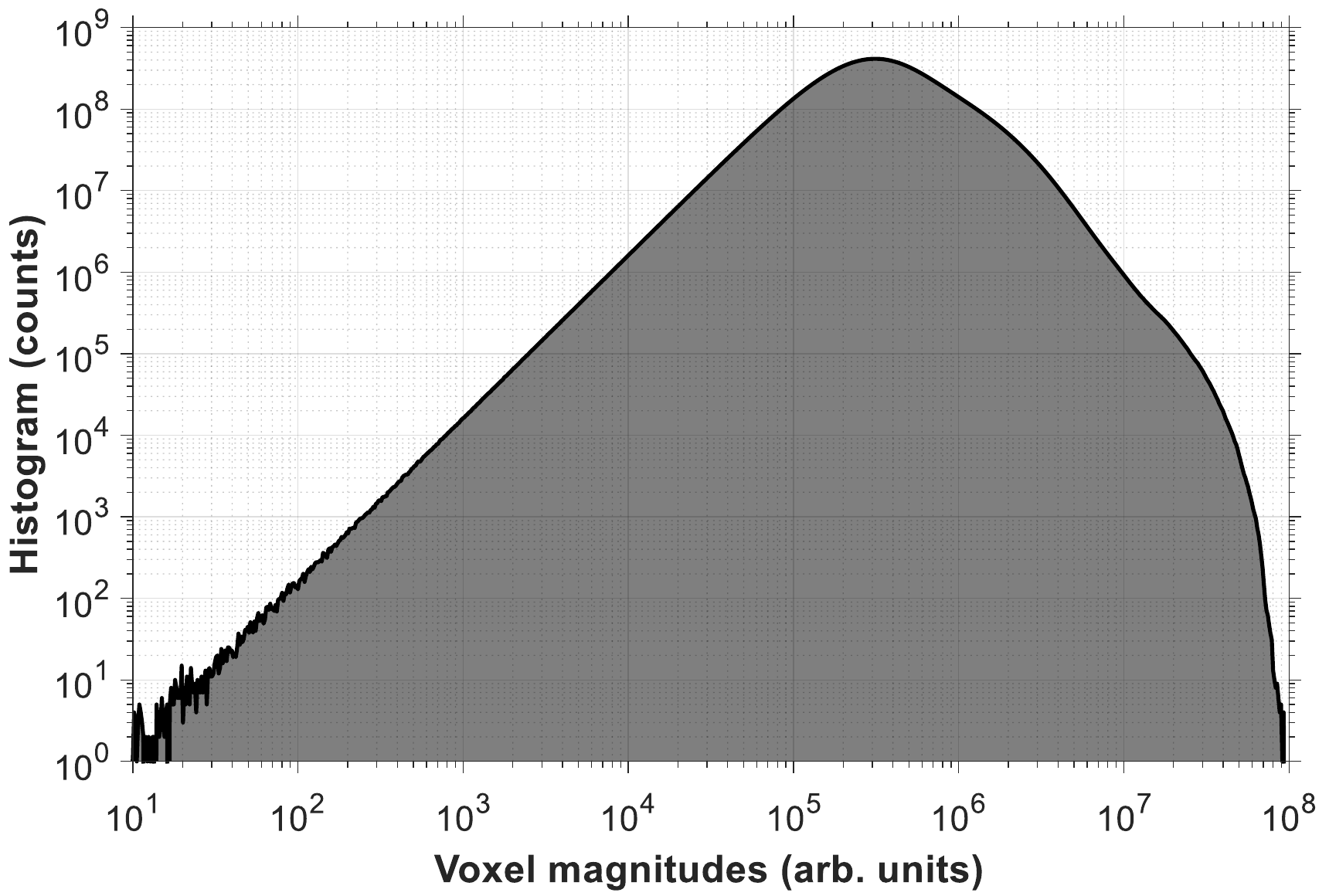}
\caption{Log-log plot of the histogram for all 45 billion voxels (with histogram bins spaced evenly in log space).}
\label{Fig4}
\end{figure}

For some intuition for the dataset, a histogram of the voxel magnitudes for all 45 billion voxels is given in Figure \ref{Fig4}. Note that this is a log-log plot. On a linear vertical axis, the distribution would simply appear bell-curve shaped, and could be interpreted to be a Gaussian distribution (that is, a lognormal distribution, since the horizontal axis is logarithmic). However, a lognormal distribution would give a symmetric parabola on a log-log plot, and it is clear to see in Figure 4 that the distribution is skewed right, with the left tail decidedly linear, not parabolic. This is in good agreement with a Weibull distribution, where in the $X\ll\lambda$ limit, the distribution should appear linear on a log-log plot. The right tail is where buried objects exist, and it is not obvious from this plot where the transition from buried object to clutter occurs. Of course, as with any detection problem, these distributions overlap. However, due to the low prevalence of buried objects, only the extreme right edge of the plot would be contaminated with buried objects. The additional structure to the plot beyond a magnitude of $\sim$10$^6$ is likely due to the strong spatial relationship the scale parameter has with space, or in particular, depth.

\begin{figure}
\centering
\subfloat[]{\includegraphics[width=0.8\linewidth]{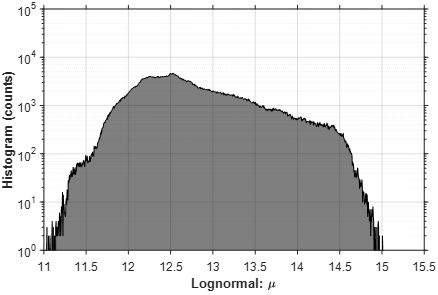} \label{Fig5a}}
\hfil
\subfloat[]{\includegraphics[width=0.8\linewidth]{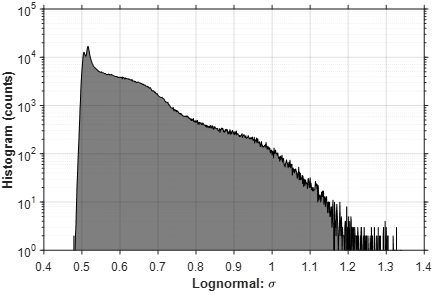} \label{Fig5b}}
\hfil
\subfloat[]{\includegraphics[width=0.8\linewidth]{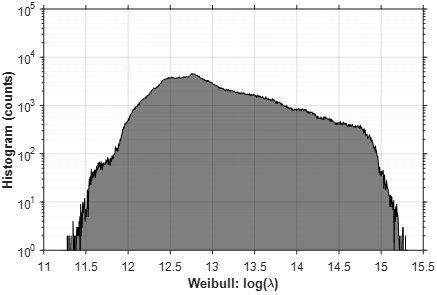} \label{Fig5c}}
\hfil
\subfloat[]{\includegraphics[width=0.8\linewidth]{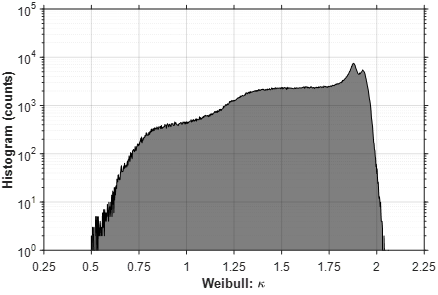} \label{Fig5d}}
\caption{Histogram of best-fit parameters for the lognormal distribution ($\mu$ and $\sigma$ are Fig. \ref{Fig5a} and \ref{Fig5b} respectively), and the Weibull distribution ($\sigma$ and $\kappa$ are Fig. \ref{Fig5c} and \ref{Fig5d} respectively)}
\label{Fig5}
\end{figure}

For the purposes of evaluating the two fit parameters for both distributions, each run, cross-track, and depth position are treated as statistically independent of each other. In other words, there are 1.1 million different shape and scale parameters for both distributions (189 runs $\times$ 128 cross-track positions $\times$ 46 depth positions = $\sim$1.1 million). A histogram of these four parameters is given in Figure \ref{Fig5}.

\subsection{Spatial variation of distribution parameters}
\begin{figure*}
\centering
\subfloat[]{\includegraphics[width=0.49\linewidth]{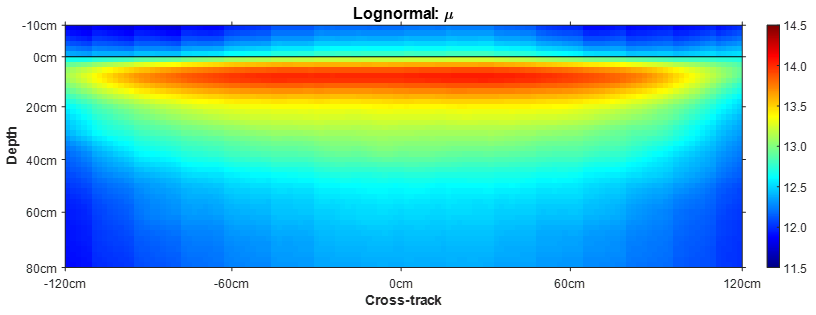} \label{Fig6c}}
\subfloat[]{\includegraphics[width=0.49\linewidth]{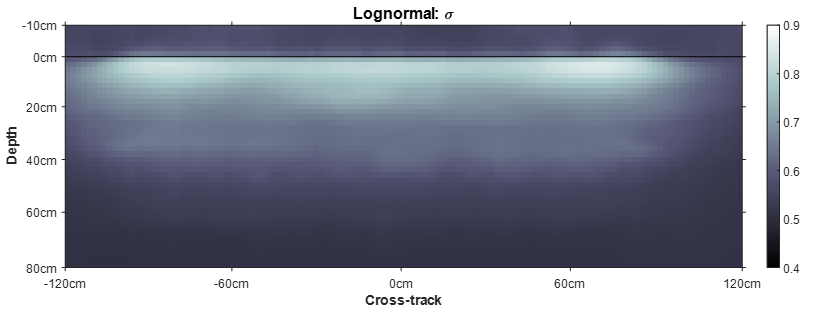} \label{Fig6d}}
\hfil
\subfloat[]{\includegraphics[width=0.49\linewidth]{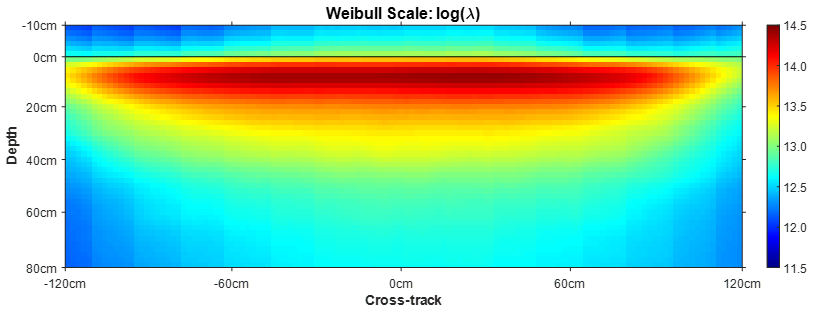} \label{Fig6e}}
\subfloat[]{\includegraphics[width=0.49\linewidth]{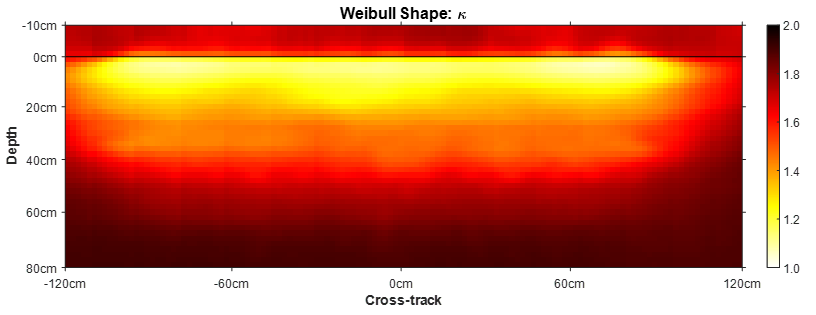} \label{Fig6f}}
\caption{Cross-track and depth variation of various moments and parameters. The left column shows scale parameters, with the top being the fitted $\mu(x,z)$ parameter for the lognormal distribution, and the bottom being the logarithm of the fitted $\lambda(x,z)$ parameter for the Weibull distribution. The right column gives the shape parameters, with the top being $\sigma(x,z)$ for the lognormal distribution, and the bottom being $\kappa(x,z)$ for the Weibull distribution. All four plots have a horizontal axis of 2.4m and a vertical axis of 0.9m, with a black horizontal line indicating the ground surface. The color axes are indicated to the right of each plot.}
\label{Fig6}
\end{figure*}

The histogram parameter plots in Fig. \ref{Fig5} hide the spatial variation of the scale and shape parameters. To illustrate those, Figure \ref{Fig6} shows the cross-track and depth variation of the four distribution parameters’ values averaged across the 189 runs. In all six plots, the values are weighted by the number of frames in each run (so that longer runs are weighted more highly than shorter runs). The possible physical interpretations for the results in Fig. \ref{Fig6} are reserved for Section \ref{Sec3}.

\subsection{Goodness of fit}
\label{Sec2_4}
\begin{figure}
    \centering
    \subfloat[]{\includegraphics[width=\linewidth]{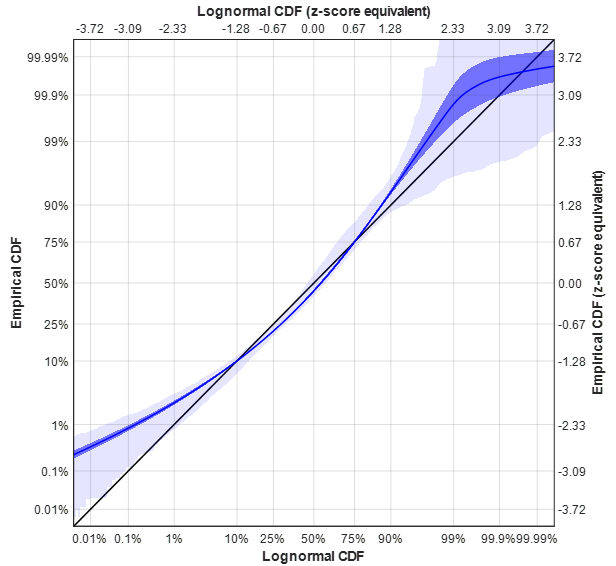} \label{Fig7a}}
    \hfil
    \subfloat[]{\includegraphics[width=\linewidth]{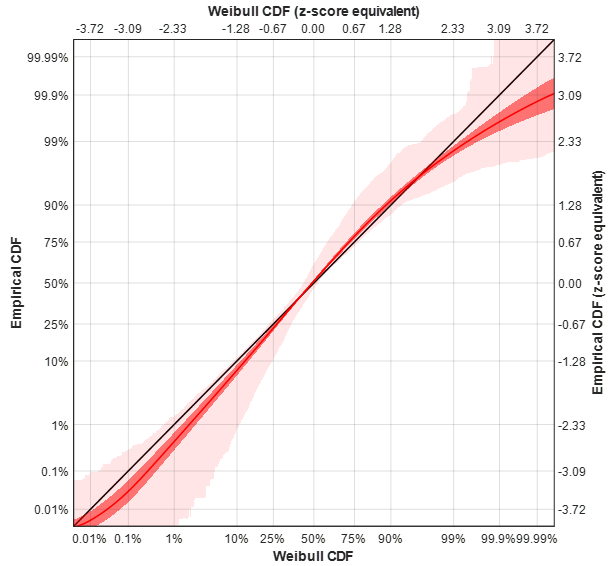} \label{Fig7b}}
    \caption{A modified Q-Q plot that allows the determination of which distribution fits the data best, with the Lognormal distribution in \ref{Fig7a}, and Weibull distribution in \ref{Fig7b}. The black diagonal line is the line of perfect match between the empirical and theoretical distributions. The blue/red lines indicate the average Q-Q plot across all runs and $(x,z)$ positions. The darker-shaded regions represent $\pm1$ standard deviation across all runs and $(x,z)$ positions. The lighter-shaded regions represent the maximum and minimum values observed. Note that the plots are shown with linear sampling in the $Z$-score equivalent axes, labeled across the top and right, though the corresponding CDF percentages are labeled on the left and bottom axes.}
    \label{Fig7}
\end{figure}

The final consideration that should be made here is the determination of which distribution fits the data best. To address that, a modified version of a Q-Q (quantile-quantile) plot is created, as shown in Figure \ref{Fig7}. To generate Fig. \ref{Fig7}, a set of linearly spaced $Z$-score values are chosen, which can be converted into their CDF equivalents by way of $1+\text{erf}\left(\sqrt{2} Z\right)=2\cdot\text{CDF}$, which is the relationship between $Z$-score and CDF for a normal distribution. Then, using each run and $(x,z)$ position’s best fit distribution parameters, those CDFs are mapped to a clutter magnitude. Next, the empirical distribution is estimated for each clutter magnitude—or in other words, what percent of the data for that run and $(x,z)$ position is less than or equal to the given clutter magnitude. This empirical CDF is evaluated for every clutter magnitude (one for each $Z$-score being sampled). Then a parametric plot of the empirical CDF vs the theoretical CDF is created for each run and $(x,z)$ position, which is effectively a Q-Q plot.

If the horizontal and vertical axes scaled linearly with CDF, then the rare (infrequent) values that help determine the fit of the distribution would be pushed into the extreme edges of the plot, with most of the plot showing strong correspondence between median values of distribution. However, by scaling linearly with $Z$-score, as these plots do, these rare values are given equal significance in the plot as the common values near the median. In other words, the middle 80\% is given the same space on the horizontal (and vertical) axis as the lower 10\% and the upper 10\%. Similarly, the 99\% to 99.9\% (or 1\% to 0.1\%) of the data takes up approximately the same space on the plot as one of the middle quartiles (\textit{e.g.} 25\% to 50\%).

This plot can be generated for all 1.1 million data sets [one for each of the runs, and $(x,z)$ positions]. To avoid plotting 1.1 million lines, instead the mean (solid colored lines), the $\pm1$ standard deviation (darker-shaded regions) and the maximum/minimum (lighter-shaded regions) are shown. The closer the lines and shaded regions are to the black diagonal line, the better the fit. If the shaded regions fall above the black line, this means the observed distribution has a fatter tail than the proposed distribution. Similarly, falling below the black line means the observed distribution has a thinner tail than the proposed distribution. Additionally, to quantify the goodness of fit, the Kolmogorov-Smirnov (KS) statistic is generated, which is effectively a measure of the largest difference between the empirical CDF and the test distribution's CDF for any quantile \cite{Drew2000}, where a smaller KS statistic implies a better fit to the proposed distribution.

These Q-Q plots support the following conclusions: 
\begin{enumerate}
    \item  The observed clutter distributions have consistently fatter tails on the left (low clutter magnitudes) than would be expected for a lognormal distribution.
    \item The observed clutter distribution has thinner tails on the left (low clutter magnitudes) than would be expected for a Weibull distribution, but the match is quantitatively better than that of the lognormal distribution.
    \item Both plots show significant deviations above the 99\% CDF. However, this is where buried objects are expected to reside, and therefore a mismatch in that region is not surprising, as image magnitudes for buried objects would be expected to followed a different distribution than for clutter, and would likely be strongly dependent on the buried objects’ scattering physics.
    \item The average KS statistic (averaged over each run and each cross-track and depth position) for the Weibull distribution (0.0312) is about half the value of the KS statistic using the Lognormal distribution (0.0555), which implies that a Weibull distribution is a better fit to the data than Lognormal.
\end{enumerate}

Additionally, it can be seen qualitatively that the Weibull plot in Fig. \ref{Fig7a} is closer to linear (particularly below the 99th percentile) than the Lognormal plot in Fig. \ref{Fig7b}, which has a consistent curve away from the black line. For all of these reasons, it is claimed that the Weibull PDF is a better fit to the observed distribution than the Lognormal PDF. Though, it is also important to point out, that for the middle (which represents $\sim 80$\% of the observed distribution), both distributions fit reasonably well. Only at the more extreme values in the left tail are deviations appreciable.

\section{Parameter Interpretations}
\label{Sec3}
This section considers the spatial variation of the distribution parameters given in the previous section. More specifically, it delineates what the spatial variation means in physical terms about either the array or the environment.

\subsection{Attenuation coefficient}
\begin{figure}
    \centering
    \includegraphics[width=0.75\linewidth]{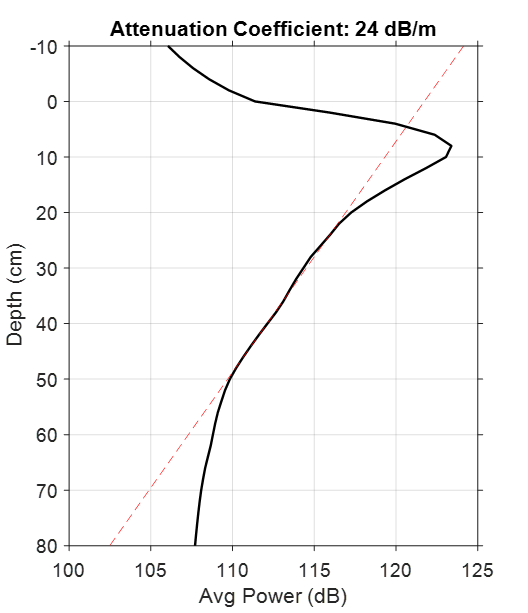}
    \caption{The average power, in decibels (dB), averaged cross-track, versus depth, in centimeters. The decibel scale is relative -- no milliwatt reference is implied in this figure. The red dashed line is a line of best fit between 20 – 50 cm of depth, which has a slope of approximately 24 dB/m, which can be interpreted as an indirect measure of the attenuation coefficient for this soil.}
    \label{Fig8}
\end{figure}

A plot of the mean linear amplitude as a function of depth is given in Figure \ref{Fig8}. The linear amplitude is averaged across all runs (weighted by the number of frames in a run), and then averaged in cross-track.

Theoretically, this plot of power vs. depth would give amplitudes near the noise floor above the surface, and then would decay linearly with increasing depth until returning to the noise floor. Roughly, that trend does hold, but only approximately so. Above the surface, the amplitudes here are likely above the would-be noise floor, as this part of the image corresponds to early arrival times, which is where the coupling pulse dominates the signal. Imperfect removal of the coupling pulse can lead to higher amplitudes in this air gap region.

The first 20 cm of soil shows an unexpected peak in amplitude. The most likely culprit for this is imperfect removal of the ground bounce, the removal of which is known to be difficult on non-flat ground, as well as the imperfectly impulsive source waveform (see Fig. \ref{Fig3}, which shows some ringing after the main pulse). However, from about 20 -- 50 cm in depth, the plot is roughly linear, with a best-fit slope of 24 dB/m, which can be considered to be an indirect measurement of the attenuation coefficient. This estimate ignores other forms of losses, such as spherical spreading, or the beam pattern of the antennas, so 24 dB/m could be considered an upper bound on the attenuation coefficient. Nonetheless, this value is reasonable, given the the frequency content of the pulse and the geology of the test location (a hot desert region whose soil type is primarily loamy sand, containing 2\% or less of silt or clay).

\subsection{Cross-track beam pattern}
\begin{figure}
    \centering
    \includegraphics[width=\linewidth]{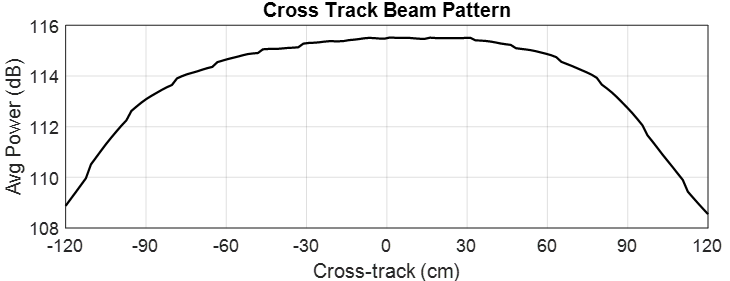}
    \caption{Average power versus cross-track, averaged across runs and in depth}
    \label{Fig9}
\end{figure}

In Figure \ref{Fig9}, the mean amplitude is given versus cross-track position, where the mean is a weighted average across runs, and averaged in depth. The small stair-steps visible in this plot are reflected in the vertical striations seen in the left column of Fig. \ref{Fig6}, and are associated with a discrete change in the number of transmitters/receivers contributing to a given voxel (\textit{i.e.} imposing the multistatic degree with a hard cut-off, rather than a smooth transition). Despite these small stair-steps, cross-track trends can be seen, which can be attributed to the cross-track beam pattern of the antennas. This is again an indirect measurement, but it shows that voxels on the extreme edges are ``illuminated'' at about half the intensity (6 dB of power) as an equivalent voxel at the center. There are likely two contributions to the cross-track amplitude variations: (1) the beam pattern of the individual antennas, and (2) the use of multi-static data with a finite aperture.

\subsection{Transition from clutter-limited to noise-limited}
\begin{figure}
    \centering
    \includegraphics[width=0.75\linewidth]{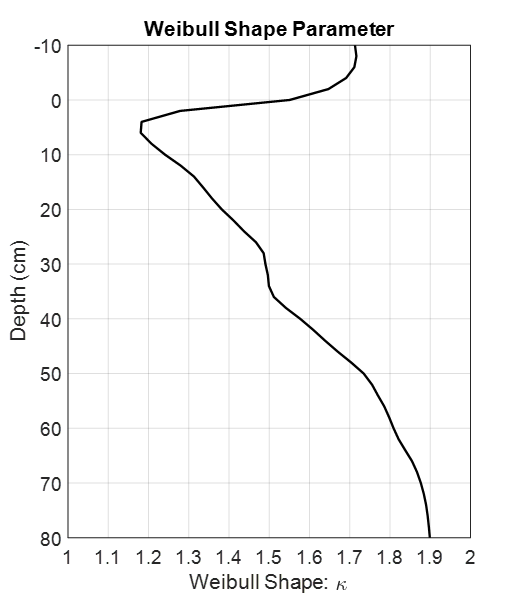}
    \caption{Plot of the average Weibull shape parameter versus depth, where the shape parameter is a weighted average across runs, and an average in cross-track. $\kappa=1$ corresponds to an exponential distribution, and $\kappa=2$ corresponds to a Rayleigh distribution.}
    \label{Fig10}
\end{figure}

Consider a plot of the average Weibull shape parameter as a function of depth, as shown in Figure \ref{Fig10}. A value of $\kappa=1$ corresponds to an exponential distribution, whereas a value of $\kappa=2$ corresponds to a Rayleigh distribution. A Rayleigh distribution is interesting because that would be the distribution one would anticipate for the magnitude of a complex number whose real and imaginary parts are zero-mean Gaussian distributed. The expectation is that noise, \textit{i.e.} thermal noise, would produce this zero-mean Gaussian distribution for the real and imaginary parts, leading to $\kappa=2$ for the best-fit Weibull distribution. Thus, for deeper voxels, the best-fit Weibull distribution is getting closer to a Rayleigh distribution, which would be the noise-limited case. However, it is not possible to claim the exact depth at which the voxel magnitudes are no longer clutter-limited, but instead noise-limited. For example, $\kappa=1.9$ might truly still be clutter-limited, though it cannot be confirmed without conducting simulations or more detailed analyses of this data. But if one uses a nominal cut-off of $\kappa=1.5$ (being halfway between the extreme cases of a Rayleigh distribution and an exponential distribution), then a penetration depth of approximately 35 cm can be estimated. Note: this does not mean targets cannot be detected deeper than this depth. What it does suggest is that a more powerful transmitter (or lower-noise receiver) would likely push this transition point deeper. Or, alternatively, if the region of interest is no deeper than 35 cm, a more powerful transmitter would not be expected to improve performance.

\section{GPR Image Standardization}
\label{Sec4}
Given knowledge of the underlying clutter distributions and their spatial variation, this information can be used to transform voxel magnitudes from their raw values into another value which more accurately reflects their deviation from the background. To do this, the CDF for each voxel can be estimated based on the underlying distribution. Plotting CDFs directly, however, is inconvenient, as the most interesting data points would occur in the range between 0.9 and 1.0. However, calculating a $Z$-score on logarithmic data allows brightness values to be compared more intuitively, on a `linear' axis, much like the rescaled axes in Fig. \ref{Fig7}. This transformed image is termed as the ``standardized logarithmic intensity'' image, or SLI image. This is the motivation behind image standardization -- the exact implementation of which is considered in the next subsection.

\subsection{Standardized Logarithmic Intensity (SLI) definition}
\label{SLIdefSection}
Consider the raw tomographic image output $R(x,y,z)$. The standardized logarithmic intensity (SLI) image output is given by:

\begin{equation}
    \text{SLI}(x,y,z)=\frac{\log R(x,y,z) - \mu(x,z)}{\sigma_{\text{avg}}}+f_{\text{airgap}}(z)
\end{equation}

The logarithm shown can be whichever base, as long as the $\mu$ and $\sigma$ used are calculated on that same base (the plots in Fig. \ref{Fig6} are generated with base $e$). $\mu(x,z)$ is the lognormal mean, which is a function of cross-track and depth, as given in Fig. \ref{Fig6c}. $\sigma_{\text{avg}}$ is in fact a scalar, and is the mean of Fig. \ref{Fig6d} (averaged in quadrature in depth and cross-track). Note that, for real-time systems, $\mu$ and $\sigma$ can be updated for each down-track position, and thus could be considered functions of $y$ as well. Finally, $f_{\text{airgap}}(z)$ is defined as $-2|z|/(10\text{cm})$ for $z<0$, and zero for $z>0$. This ad-hoc function was designed to suppress air gap imaging artifacts arising from pre-processing that are unintentionally accentuated in image standardization. While this function is designed to suppress voxels in the airgap, it is also designed to have no effect on the subsurface voxels, which is where the most interesting image features occur.

The choice in the above equation to use $\sigma_{\text{avg}}$ rather than $\sigma(x,z)$ is somewhat arbitrary, but has its advantages. First, in a real-time implementation of this system, it is possible that $\sigma(x,z)$ may take many frames to converge to an appropriate value, whereas $\sigma_{\text{avg}}$ would be expected to converge somewhat faster. Additionally, $\sigma(x,z)$ was found to be a weak function of spatial position (see Fig. \ref{Fig6d}), so the approximation that $\sigma(x,z)\approx\sigma_{\text{avg}}$ is reasonable. Finally, it also preserves the interpretation of $\text{SLI}(x,y,z)$ being approximately equal  to a normalization of $R(x,y,z)$ instead of a standardization (where normalization refers to dividing by a constant, and standardization refers to subtracting a mean and dividing by a standard deviation). Owing to the nature of logarithms, and ignoring the $f_\text{airgap}(z)$ function above (which is zero for subsurface voxels anyway), $\text{SLI}(x,y,z)$ can be re-written as:

\begin{equation}
    \text{SLI}(x,y,z) = \log\left(\left[\frac{R(x,y,z)}{\exp(\mu(x,z))}\right]^{1/\sigma_\text{avg}}\right)
\end{equation}

In this form, it is more easily seen that SLI images are simply logarithms of the normalized image, which is then rescaled by an exponent of $1/\sigma_\text{avg}$. Thus, even though SLI is named as a standardized image, and its definition certainly resembles a standardization process, it can also be thought of as substantively equivalent to a normalized image, thanks to the use of a constant $\sigma_\text{avg}$, as opposed to a spatially varying one. Additionally, keeping $\sigma$ constant means only $\mu(x,z)$ can amplify or attenuate data, not $\sigma$ as well.

Also, it should be noted that the SLI definition utilizes magnitudes, not intensities, though its name contains the word ``intensity''. However, due to the logarithm, using true intensities (proportional to magnitude-squared) would produce a $\mu$ and $\sigma$ that are a factor of two larger, and therefore the standardization equation would cancel out this effect.

\subsection{Standardized images of targets}
\begin{figure*}
    \centering
    \subfloat[]{\includegraphics[width=0.45\linewidth]{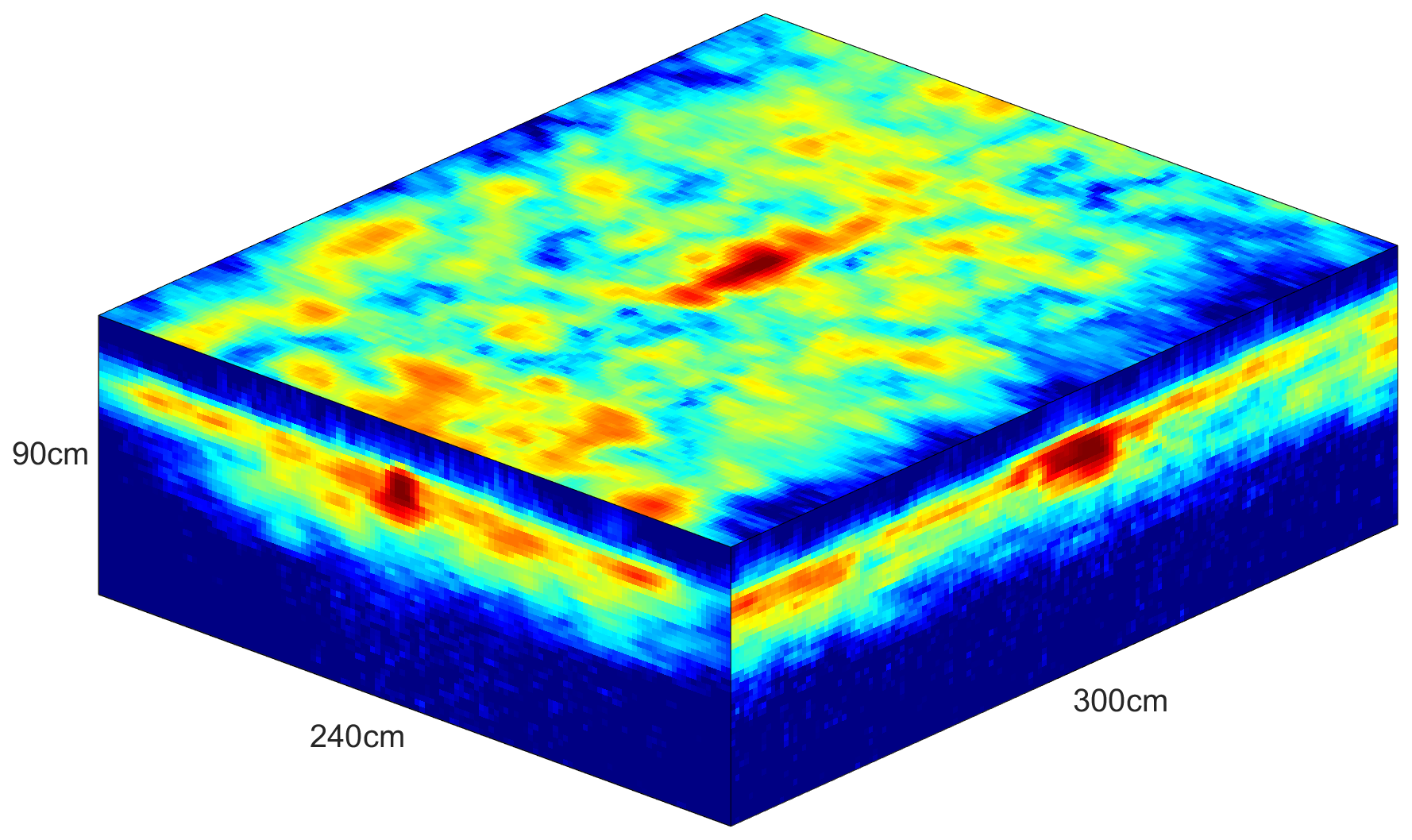}}
    \subfloat[]{\includegraphics[width=0.45\linewidth]{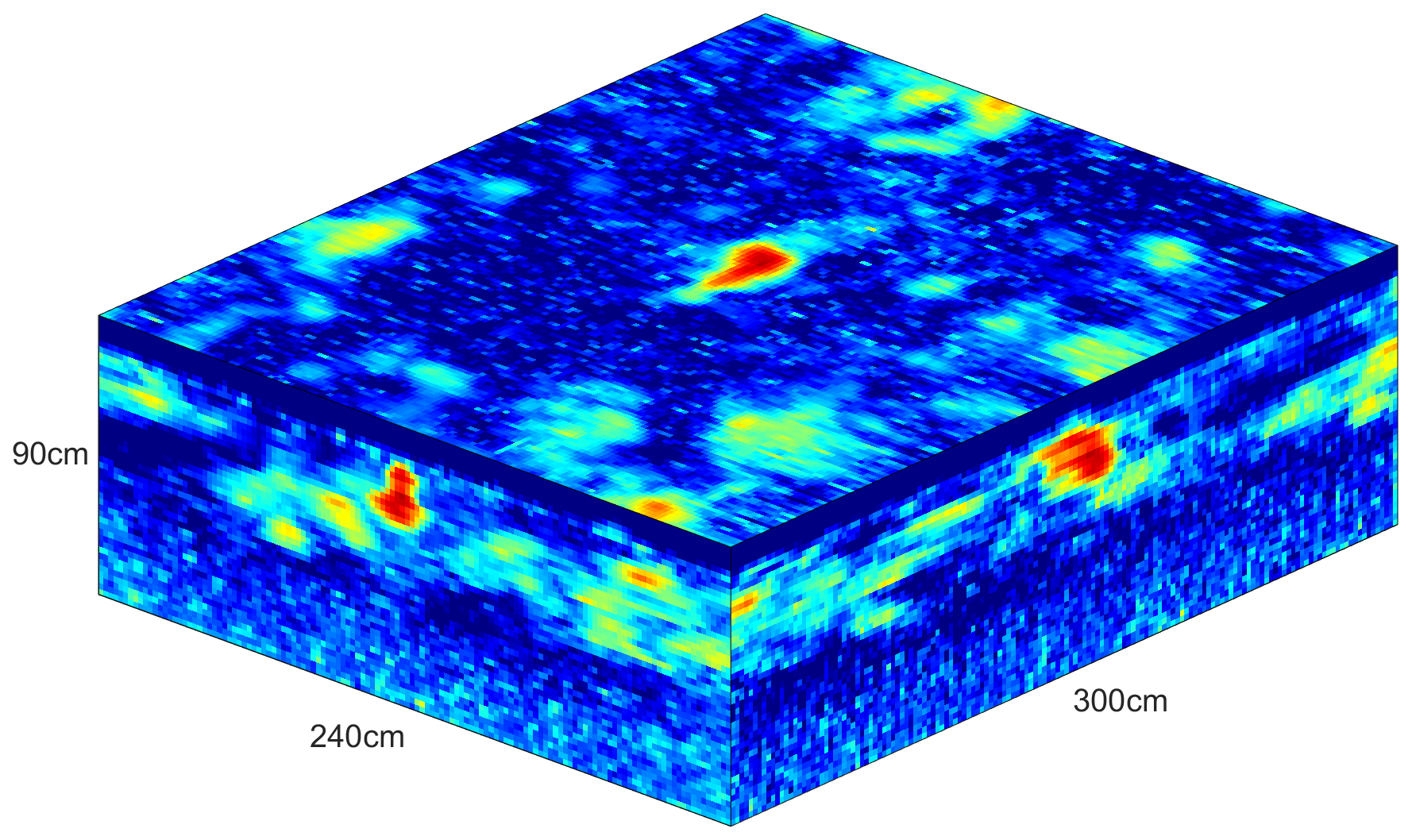}}
    \hfil
    \subfloat[]{\includegraphics[width=0.45\linewidth]{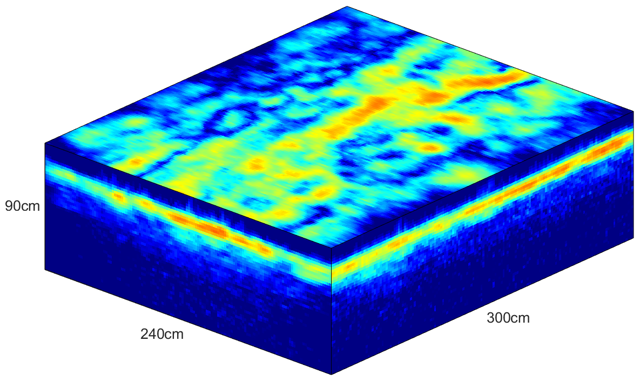}}
    \subfloat[]{\includegraphics[width=0.45\linewidth]{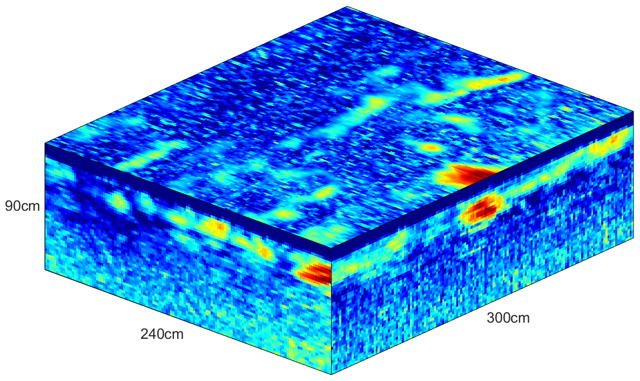}}
    \hfil
    \subfloat[]{\includegraphics[width=0.45\linewidth]{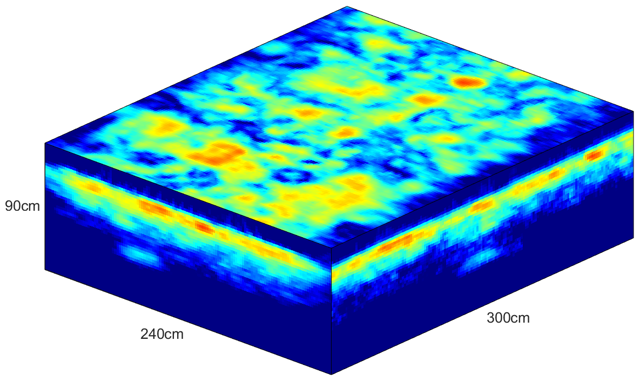}}
    \subfloat[]{\includegraphics[width=0.45\linewidth]{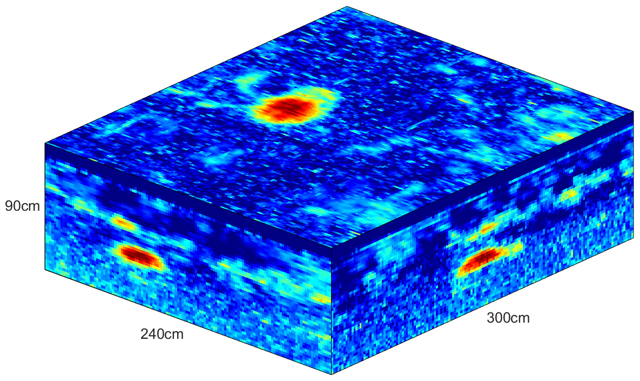}}
    \hfil
    \subfloat[]{\includegraphics[width=0.45\linewidth]{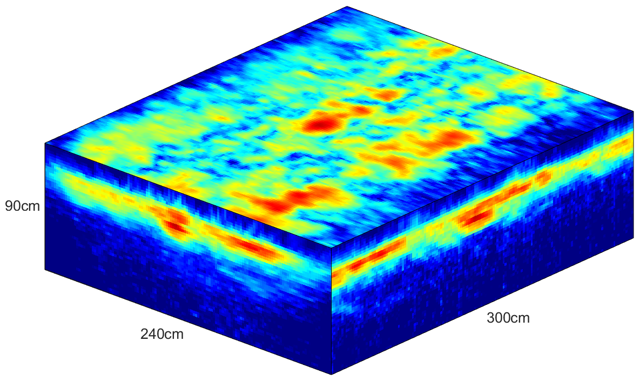}}
    \subfloat[]{\includegraphics[width=0.45\linewidth]{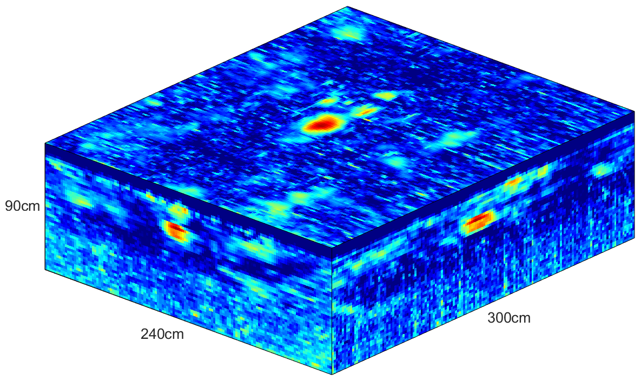}}
    \caption{Raw images (left column) and standardized images (right column) for four different targets: a circular plastic-encased landmine (first row), a large, square, plastic landmine seen at the edge of the array (second row), a deep large, metallic target (third row), and a shallow plastic-encased circular landmine surrounded by bright clutter (fourth row).}
    \label{Fig11}
\end{figure*}

Four example targets are selected to illustrate the utility of image standardization. In Figure \ref{Fig11}, four pairs of images are shown. On the left are raw images, shown after a logarithm is applied, i.e. $\text{log}(R(x,y,z))$, and on the right are $\text{SLI}(x,y,z)$ images, standardized according to Section \ref{SLIdefSection}. In all four raw image plots, the raw color scale is defined to be between 13 (dark blue) and 16 (dark red). In all four standardized image plots, the SLI color scale is set to be between 0.5 (dark blue) and 2.5 (dark red). Since the images are fundamentally three-dimensional volumes, a max projection across each of the three axes is used to display these cubes. Recall that $x$ (cross-track) is the 240-cm-long dimension, $y$ (along-track) is the 300-cm-long dimension, and $z$ (depth) is the 90-cm-long dimension. The upper dark blue region seen in the $x$--$z$ and $y$--$z$ plots show the air gap, where the region is made darker blue than expected due to the use of $f_\text{airgap}(z)$ in the SLI definition.

For these targets, the standardized images are generally superior when detecting buried objects, as the target can technically be seen in the raw images in all four cases, but the presence of shallow clutter tends to obscure them (here, `shallow' refers to depths of a few wavelengths). By standardizing with respect to the underlying clutter distribution, only voxels with anomalously high magnitudes are accentuated, with the rest being suppressed. These four images illustrate that standardization is generally positive for improving the contrast between clutter and targets, which is convenient for the development of an automated detection algorithm. The target in the first row is fairly typical, in that the target can be seen in unstandardized image, but the standardized image (Fig \ref{Fig11}b) makes it clearer. The other three target examples were chosen to accentuate a different aspect of the standardization: compensating for cross-track variation (Fig \ref{Fig11}d), compensating for depth variation (Fig \ref{Fig11}f), and the suppression of shallow clutter to reveal a slightly deeper target (Fig \ref{Fig11}h).

While these figures show an improvement in contrast with standardization, note that this apparent improvement in contrast is primarily associated with compensating for the spatial variation of the clutter distribution. Standardization by itself does not increase intrinsic contrast, it simply measures contrast in a way that can be easily compared with other regions of the image, or with other systems.

Standardizing in this way relies upon the underlying clutter distribution being unimodal for a given $(x,z)$ row, with well-defined means and variances. The unimodal requirement is typically satisfied, unless there are significant along-track variations in the environment that could lead to multimodal distributions. The means and variances of multimodal distributions are less meaningful for standardization, and could present challenges. However, if multimodal distribution occurred due to the background environment changing discontinuously (ie. the vehicle abruptly transitioned from soil to concrete), the running mean and variance calculation could be reset based on some threshold. Other than this, only pathological clutter distributions, such as the Cauchy-Lorentz distribution whose variance is undefined, would present difficulties for this method of standardization. In the authors' experience, this methodology to standardize GPR images was found to be robust across a variety of testing locations and GPR imaging settings.

\section{Conclusions}
\label{Sec5}
This study supports the following three conclusions:
\begin{enumerate}
\item The underlying clutter magnitude distribution in GPR imagery is modeled well by a Weibull distribution rather than a log-normal distribution. This suggests that clutter magnitudes in GPR imagery follows similar distributions as sea-clutter gathered from synthetic aperture radars.
\item The Weibull shape parameter can be used to estimate the transition in depth from clutter-limited to noise-limited. In this study, objects at a depth of 35 cm or less are unlikely to be improved by a more powerful transmitter, whereas deeper objects could benefit from a more powerful transmitter.
\item Image standardization that appropriately compensates for attenuation and finite antenna beam-widths leads to images with significantly improved clutter-to-target contrast, and allows for more fair comparison between targets seen at the edge of the array, deep targets, and shallow targets surrounded by bright clutter.
\end{enumerate}

\section*{Acknowledgment}

The authors would like to acknowledge support from the Office of Naval Research.
This work was performed under the auspices of the U.S. Department of Energy by Lawrence Livermore National Laboratory under Contract DE-AC52-07NA27344.

% Can use something like this to put references on a page
% by themselves when using endfloat and the captionsoff option.
\ifCLASSOPTIONcaptionsoff
  \newpage
\fi

% trigger a \newpage just before the given reference
% number - used to balance the columns on the last page
% adjust value as needed - may need to be readjusted if
% the document is modified later
%\IEEEtriggeratref{8}
% The "triggered" command can be changed if desired:
%\IEEEtriggercmd{\enlargethispage{-5in}}

% references section

% can use a bibliography generated by BibTeX as a .bbl file
% BibTeX documentation can be easily obtained at:
% http://mirror.ctan.org/biblio/bibtex/contrib/doc/
% The IEEEtran BibTeX style support page is at:
% http://www.michaelshell.org/tex/ieeetran/bibtex/
\bibliographystyle{IEEEtran}
% argument is your BibTeX string definitions and bibliography database(s)
\bibliography{IEEEabrv,references}

%
% <OR> manually copy in the resultant .bbl file
% set second argument of \begin to the number of references
% (used to reserve space for the reference number labels box)
% \begin{thebibliography}{1}
%
%\bibitem{IEEEhowto:kopka}
%H.~Kopka and P.~W. Daly, \emph{A Guide to \LaTeX}, 3rd~ed.\hskip 1em plus
%  0.5em minus 0.4em\relax Harlow, England: Addison-Wesley, 1999.%
%
%\end{thebibliography}

% biography section
% 
% If you have an EPS/PDF photo (graphicx package needed) extra braces are
% needed around the contents of the optional argument to biography to prevent
% the LaTeX parser from getting confused when it sees the complicated
% \includegraphics command within an optional argument. (You could create
% your own custom macro containing the \includegraphics command to make things
% simpler here.)
%\begin{IEEEbiography}[{\includegraphics[width=1in,height=1.25in,clip,keepaspectratio]{mshell}}]{Michael Shell}
% or if you just want to reserve a space for a photo:

\begin{IEEEbiography}[{\includegraphics[width=1in,height=1.25in,clip,keepaspectratio]{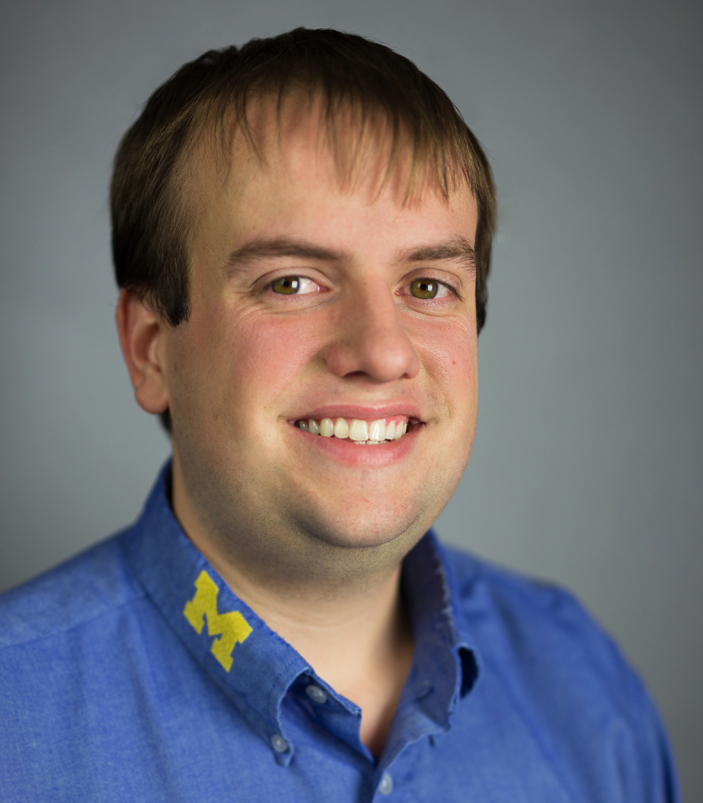}}]{Brian M. Worthmann}
is a postdoctoral researcher at LLNL with a background in wave-based remote sensing and signal processing. He received his PhD at the University of Michigan in Applied Physics, where he worked on model-based source localization in ocean acoustics. At LLNL, he has worked with ground penetrating radar signal and image processing, as well as statistical data analysis for hardware and performance characterization.
\end{IEEEbiography}

\begin{IEEEbiography}[{\includegraphics[width=1in,height=1.25in,clip,keepaspectratio]{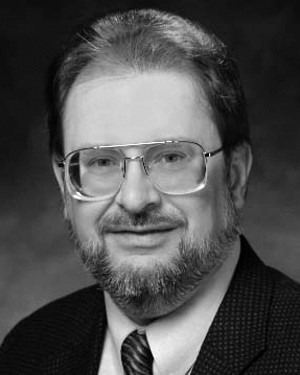}}]{David H. Chambers}
David H. Chambers (M’97–SM’04) is originally from southeast Kansas. He received the Bachelor’s degrees in both physics and mechanical engineering, and the Master’s degree in physics from Washington University in St. Louis, St Louis, MO, USA, from 1976 to 1982, and the Ph.D. degree in theoretical and applied mechanics from the University of Illinois at Urbana-Champaign, in 1987. Afterward, he took a position as a Physicist with Lawrence Livermore National Laboratory, Livermore, CA, USA, working on the propagation of high-energy laser beams through the atmosphere and radar imaging of the ocean surface. His research interests include ground penetrating radar, gravity gradiometry, statistical theory of fission chains, and applications of time reversal symmetry to target characterization and communications for both acoustic and radar imaging. Dr. Chambers is a Fellow of the Acoustical Society of America and a member of the American Physical Society and Society of Industrial and Applied Mathematics. He has published papers in the IEEE Transactions for Antennas and Propagation, Journal of the Acoustical Society of America, Physical Review, and others.
\end{IEEEbiography}

\begin{IEEEbiography}[{\includegraphics[width=1in,height=1.25in,clip,keepaspectratio]{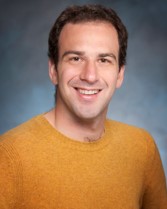}}]{David S. Perlmutter}
David S. Perlmutter is a staff engineer at Lawrence Livermore National Laboratory (LLNL) interested in applied signal processing and machine learning. He received a M.S. in electrical engineering at the University of Washington (’15), where he researched techniques for low-dose CT image reconstruction. Previously he was a staff engineer at MIT Lincoln Laboratory, working on LiDAR imaging technologies. At LLNL, his experience includes modeling, control and data analysis of ultrafast optical and RF sensing systems.
\end{IEEEbiography}

\begin{IEEEbiography}[{\includegraphics[width=1in,height=1.25in,clip,keepaspectratio]{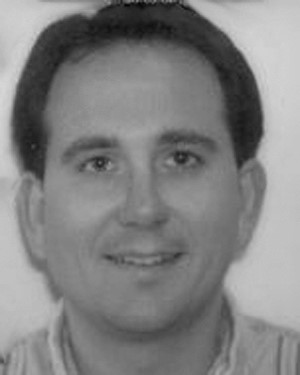}}]{Jeffrey E. Mast}
Jeffrey E. Mast received the B.S., M.S., and Ph.D. degrees in electrical and computer engineering from the University of Illinois at Urbana-Champaign, Champaign, IL, USA, in 1987, 1989, and 1993, respectively. He was a General Motors Scholar as an undergraduate and graduated with honors. As a graduate student, he was a Fellow for the U.S. Army Advanced Construction Technology Center specializing in radar imaging systems for nondestructive evaluation of civil structures. From 1993 to 1998, he held a position at Lawrence Livermore National Laboratory (LLNL) in the Imaging and Detection Program developing radar imaging systems for submarine detection, bridge deck and roadway inspection, and unexploded ordnance detection. He is currently President of Teres Technologies, Inc., Loveland, CO, USA, and serves as a Consultant for LLNL and several private and public companies. His research interests include radar imaging and remote sensing, wide area motion imagery, high performance computing, and real-time systems.
\end{IEEEbiography}

\begin{IEEEbiography}[{\includegraphics[width=1in,height=1.25in,clip,keepaspectratio]{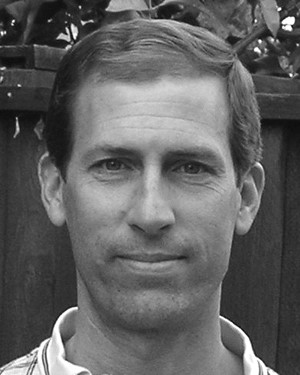}}]{David W. Paglieroni}
David W. Paglieroni (S’80--M’87--SM’02) received the B.S. (summa cum laude), M.S., and Ph.D. degrees in electrical and computer engineering from the University of California, Davis, CA, USA, in 1982, 1984, and 1986. In 1987, he joined Lockheed Martin, San Jose, CA, USA, where he served as a Manager of the Imagery Software Section. Since 1999, he has been with the Lawrence Livermore National Laboratory (LLNL), Livermore, CA, USA, where he is currently a senior member of the technical staff in the Engineering Directorate.  Dr. Paglieroni provides subject matter expertise on a variety of national security projects.  He holds numerous patents and has published numerous peer-reviewed papers in various journals and conferences.  His current areas of interest and expertise include computer vision and AI-based data fusion for risk assessment / situational awareness.
\end{IEEEbiography}

\begin{IEEEbiography}[{\includegraphics[width=1in,height=1.25in,clip,keepaspectratio]{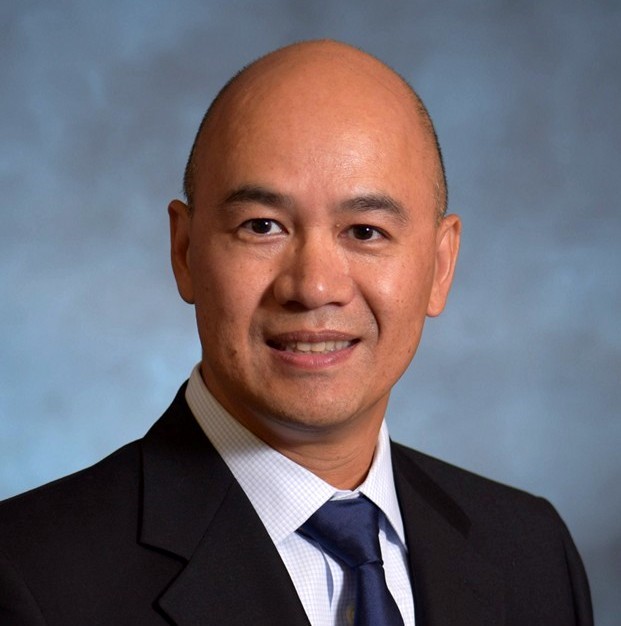}}]{Christian T. Pechard}
Christian Pechard received his M.S degree in computer engineering from ESIEE-Paris, France, in 1995. From 1996 to 2007, he worked in the field of networking, telecommunication, real-time embedded systems, network management at Cisco Systems, San Jose, California. From 2007 to 2010, he worked in the field of enterprise wireless, real-time embedded system for Trapeze Networks, Pleasanton, California. Since 2010, he joined Lawrence Livermore National Laboratory (LLNL) to research and develop applications in the field of telecommunication, wireless protocols, real-time embedded systems. He is currently responsible for software pipeline development on a large multi-years ground penetrating radar program.
\end{IEEEbiography}

\begin{IEEEbiography}[{\includegraphics[width=1in,height=1.25in,clip,keepaspectratio]{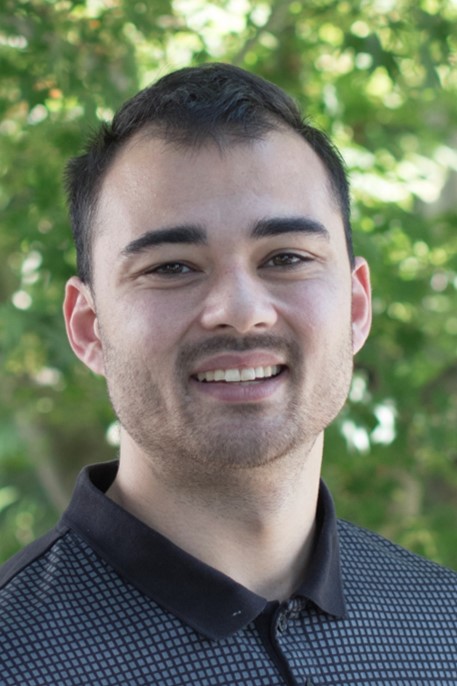}}]{Garrett A. Stevenson}
is a signal and image processing engineer at LLNL with a background in embedded deep learning, computer vision, nondestructive characterization, and navigation/localization. He received his M.S. in Computer Science at C.S.U. East Bay with emphases in Computer Vision and Embedded Systems. He currently works in drug discovery and explosives detection using CT images in Ground Penetrating Radar and Checked Baggage. 
\end{IEEEbiography}

\begin{IEEEbiography}[{\includegraphics[width=1in,height=1.25in,clip,keepaspectratio]{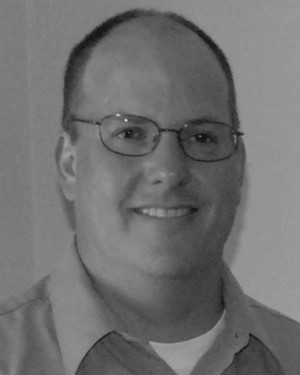}}]{Steven W. Bond}
Steven W. Bond (S’88–M’98) received the Bachelor’s (summa cum laude), M.S., and Ph.D. degrees in electrical engineering from Georgia Institute of Technology, Atlanta, GA, USA, in 1991, 1994, and 2001. Since 1998, he has been employed as a Senior Research Engineer with the National Security Engineering Division, Lawrence Livermore National Laboratory, Livermore, CA, USA. He has worked as a Photonic Device and Systems Engineer on WDM VCSEL computer interconnects, long distance (greater than 25km) free space optical communications systems, and ground penetrating radar imaging systems.
\end{IEEEbiography}

% insert where needed to balance the two columns on the last page with
% biographies
%\newpage

% You can push biographies down or up by placing
% a \vfill before or after them. The appropriate
% use of \vfill depends on what kind of text is
% on the last page and whether or not the columns
% are being equalized.

%\vfill

% Can be used to pull up biographies so that the bottom of the last one
% is flush with the other column.
%\enlargethispage{-5in}

% that's all folks
\end{document}